 \documentclass[manuscript]{emulateapj-rtx4}

\newcommand{\rhoGJ}{\rho_{{\rm GJ}}}  
\newcommand{\rlc}{\varpi_{\rm LC}} 

\slugcomment{}

\shorttitle{3-D non-vacuum outer-gap model}
\shortauthors{Hirotani}

\begin{document}


\title{Three-dimensional non-vacuum pulsar outer-gap model:
 Localized acceleration electric field in the higher altitudes}


\author{Kouichi Hirotani\altaffilmark{1}}
\affil{Academia Sinica, Institute of Astronomy and Astrophysics (ASIAA),
       PO Box 23-141, Taipei, Taiwan}

%
%
%
%


\begin{abstract}
We investigate the particle accelerator that arises
in a rotating neutron-star magnetosphere.
Solving the Poisson equation for the electro-static potential,
the Boltzmann equations for relativistic electrons and positrons,
and the radiative transfer equation simultaneously,
we demonstrate that the electric field 
is substantially screened along the magnetic field lines
by the pairs that are created and separated within the accelerator.
As a result, the magnetic-field-aligned electric field is localized 
in the higher altitudes near the light cylinder
and efficiently accelerates the positrons created in the lower altitudes
outwards but not the electrons inwards.
The resulting photon flux becomes predominantly outwards,
leading to typical double-peak light curves,
which are commonly observed from many high-energy pulsars.
\end{abstract}





\section{Introduction}
%
%
The Large Area Telescope (LAT) aboard the 
{\it Fermi Gamma-Ray Space Telescope}
has detected 117 rotation-powered pulsars \citep{abdo13}
\footnote{
For the latest Fermi discoveries, see also
https://confluence.slac.stanford.edu/display/GLAMCOG \\
/Public+List+of+LAT-Detected+Gamma-Ray+Pulsars
}.
The unprecedented amount of data for these $\gamma$-ray sources,
allows us to study the statistical properties of the
high-energy pulsars through the light-curve analysis.
Adopting the polar-cap (PC) model
\citep{stur71,harding78,daugherty82,dermer94},
the slot-gap (SG) model
\citep{arons83,musl04,dyks03,hard05}, 
and the outer gap (OG) model
\citep{cheng86a,romani96,cheng00,romani10},
and comparing the predicted light-curve morphology with the observations
\citep{dyks04},
they revealed that 
the SG model is geometrically favoured in some cases 
but the OG model is in other cases 
as opposed to the lower-altitude emission models such as the 
PC model \citep{romani10,taka11,pierb12,pierb14,johnson14}.
Moreover, MAGIC and VERITAS experiments reported
pulsed signals from the Crab pulsar up to 400~GeV
\citep{ale11a,ale11b,aliu11,aliu14},
which indicates that such very-high-energy photons
are probably emitted from the higher altitudes
to avoid the strong magnetic absorption that would arise near the 
polar-cap (PC) surface. 

%
%

In an OG, created pairs are separated by 
the magnetic-field-aligned electric field, $E_\parallel$.
For middle-aged pulsars, 
pairs are mostly created near the inner boundary;
thus, the out-going particles run
much longer distance in the OG than the in-coming ones,
resulting in an order of magnitude greater outward flux 
than the inward one (fig.~7 \& 8 of \citet{hiro02}). 
On the other hand,
for relatively young pulsars such as the Vela pulsar,
the outward $\gamma$-ray flux dominates the inward one
by only several times \citep{taka08},
because the out-going particles run the strong-$E_\parallel$ region
several times longer distance than the in-coming ones.
However, such a non-negligible inward flux 
leads to the light curve that generally exhibits more than two peaks
in a single neutron star rotation, which contradicts with observations.

Although it is unclear 
if the OG model does predict a dominant outward $\gamma$-ray flux 
for young or relatively young pulsars,
they have assumed so to obtain the observed, double-peaked light curves
\citep{roma95,cheng00,zhang02,taka07,tang08,romani10,bai10a,bai10b,vent12,
       pierb12,pierb14,johnson14}. 
We are, therefore, motivated by the need to contrive 
a more consistent OG model that quantifies the outward and inward
$\gamma$-ray fluxes,
incorporating the screening effect of 
$E_\parallel$ by the created and separated charges in the gap.
Extending the idea of \citet{bes92}, 
\citet{hiro99a,hiro99b} first solved the set of 
the non-vacuum Poisson equation, 
Boltzmann equations for electrons and positrons ($e^\pm$),
and the radiative transfer equation 
simultaneously in a pulsar magnetosphere.
It has been demonstrated \citep{taka04,hiro06b,hiro13}
that a strong $E_\parallel$ does arise 
between the null-charge surface and the light cylinder (LC),
whose distance from the rotation axis is given by the
light cylinder radius, $\varpi_{\rm LC}=c/\Omega$,
where $c$ designates the speed of light and 
$\Omega$ the rotational frequency of the neutron star.

In the present letter, we numerically solve the non-vacuum OG 
electrodynamics in the three-dimensional (3-D) magnetosphere
of a typical young pulsar
and demonstrate that the outward photon flux
naturally dominates the inward ones by virtue of the
$E_\parallel$ screening due to the separated motion of the created
$e^\pm$'s.
Without loss of any generality, we can assume a positive $E_\parallel$;
in this case, $e^+$'s are accelerated outwards while $e^-$'s inwards,
forming an outward current in the OG 
as a part of the global current circuit.
We do not solve the global current closure issue,
assuming a starward return current in the magnetic polar regions.
We define the magnetic coordinates in \S~\ref{sec:coord}
and describe the 3-D vacuum OG model in \S~\ref{sec:vac_3D}.
We then propose the new, 3-D non-vacuum OG model 
in \S~\ref{sec:nonvac_3D},
and discuss some implications of this modern OG model
in \S~\ref{sec:disc}.

\section{3-D Magnetic coordinates}
\label{sec:coord}
In a rotating magnetosphere,
the Poisson equation for the non-corotational potential $\Psi$ becomes
\begin{equation}
  -\nabla^2 \Psi = 4\pi(\rho-\rhoGJ),
  \label{eq:Poisson_1}
\end{equation}
where $\rho$ denotes the real charge density
and the Goldreich-Julian (GJ) charge density is defined by
\begin{equation}
  \rhoGJ 
  \equiv -\frac{\mbox{\boldmath$\Omega$}\cdot\mbox{\boldmath$B$}}
               {2\pi c}
         +\frac{(\mbox{\boldmath$\Omega$}\times\mbox{\boldmath$r$})\cdot
                (\nabla\times\mbox{\boldmath$B$})}
               {4\pi c}.
  \label{eq:def_rhoGJ_1}
\end{equation}
The acceleration electric field can be computed by
\begin{equation}
  E_\parallel \equiv \mbox{\boldmath$B$} \cdot \mbox{\boldmath$E$}/B
       = - \partial \Psi / \partial s,
  \label{eq:def_Ell}
\end{equation}
where $s$ denotes the distance along the magnetic field line.

To specify the position in a three-dimensional (3-D) pulsar magnetosphere,
it is convenient to introduce the magnetic coordinates 
($s$,$\theta_\ast$,$\varphi_\ast$),
where $\theta_\ast$ and $\varphi_\ast$ represent the foot point 
of the field lines on the neutron star (NS) surface.
Their relationship with the polar coordinates 
is given by equations~(15)--(17) of \citet{hiro06a}.
The magnetic azimuthal angle $\varphi_\ast$ is defined
counter-clockwise around the magnetic dipole axis;
$\varphi_\ast=0$ points the opposite direction to the rotation axis
from the magnetic axis on the two-dimensional poloidal plane
in which both the rotation and magnetic axes reside.
Thus, a negative $\varphi_\ast$ represents a magnetic field line
in the trailing side of a rotating magnetosphere,
while a positive $\varphi_\ast$ does that in the leading side.

%

As for the magnetic colatitudes $\theta_\ast$,
it is convenient to replace it with 
the dimensionless trans-field coordinate $h$ such that
\begin{equation}
  h \equiv 1-\theta_\ast/\theta_\ast^{\rm max},
  \label{eq:def_h}
\end{equation}
where $\theta_\ast^{\rm max}=\theta_\ast^{\rm max}(\varphi_\ast)$ 
describes the PC rim, 
outside of which the magnetic field lines close within the LC.
In what follows, we use the coordinates $(s,h,\varphi_\ast)$ 
to specify points in the 3-D magnetosphere. 
In an OG, $E_\parallel$ vanishes on the last-open field line, $h=0$.
In the convex side of the magnetic field lines, 
$E_\parallel$ increases nearly quadratically with increasing $h$
at each ($s$,$\varphi_\ast$),
attain the maximum value near the central height $h=0.5 h_{\rm m}$, 
then reduces to vanish above a certain height 
$h_{\rm m}=h_{\rm m}(s,\varphi_\ast)$,
which forms the upper boundary of the OG.
Here, the gap trans-field thickness $h_{\rm m}$
corresponds to $f$ in \citet{cheng86a} and 
$w$ in \citet{romani96}.
Because $E_\parallel$ is screened by the created pairs,
we obtain $h_{\rm m} \sim 0.1$ for very young pulsars like the Crab pulsar, 
while $h_{\rm m}>0.5$ for middle-aged pulsars like the Geminga pulsar.
In the non-vacuum OG model, the upper boundary $h_{\rm m}(s,\varphi_\ast)$
is consistently determined from the separating motion of
the charges by the Poisson and the Boltzmann equations.

To describe the magnetic field,
we adopt the vacuum rotating dipole solution \citep{cheng00}
in the entire simulation region.
Although this approximation breaks down near and outside the LC 
\citep{spit06}, 
it properly gives the outward/inward flux ratio
at lease qualitatively,
because the screening of $E_\parallel$ takes place within the LC.

To solve the Poisson equation and the $e^\pm$ Boltzmann equations,
we adopt $300$ bins in $s$ direction 
(from the PC surface $s=0$ to $s=3 \rlc$ along each magnetic field line),
$72$ bins in $h$ direction
(from the lower boundary $h=0$ to $h=3h_{\rm max}$), and
$96$ bins in $\varphi_\ast$ direction
(from $\varphi=-\pi$ to $\varphi=\pi$);
here, $h_{\rm max}$ refers to the maximum value of 
$h_{\rm m}(s,\varphi_\ast)$.
The outer boundary of the gap is determined 
as the free boundary at which $E_\parallel$ vanishes.
For example, if the OG is vacuum ($\rho=0$)
and thin ($h_{\rm max} \ll 1$), 
the outer boundary is located at the
inflection point where the poloidal magnetic field con-
figuration changes from convex to concave (eq. [68] of \citep{hiro06a}).
If the OG is non-vacuum or thick,
we must solve the Poisson equation to find the outer boundary.
Although there is no physical reason why an OG outer boundary
should be located within the LC
(because the LC is not a special place for any physical process), 
we impose that the maximum distance of the outer boundary 
from the rotation axis is $0.9 \rlc$
to take a consistency with classic OG models.
To solve the radiative transfer equation, 
we employ the same magnetic coordinates as the Poisson and 
$e^\pm$ Boltzmann equations with coarse grids:
$25$ bins in $s$ direction
(from $s=0$ to $s=3 \rlc$),
$10$ bins in $h$ direction 
(from the lower boundary $h=0$ to $h=2.5h_{\rm max}$, and
$96$ bins in $\varphi_\ast$ direction
(from in $\varphi_\ast=-\pi$ to $\varphi_\ast=\pi$).
In the momentum space, we adopt
$43$ bins for the photon energy 
(from $0.005$~eV to $20$~TeV),
$40$ bins for the latitudinal propagation direction
(from $0$ to $\pi$ radian)
with respect to the rotation axis, and
$60$ bins for the azimuthal propagation direction
(from $-\pi$ to $\pi$ radian).

We employ the minimal cooling scenario (Page et al. 2004), 
which has no enhanced cooling that could
result from any of the direct Urca processes and adopts
the standard equation of state, APR EOS (Akmal et al. 1998).
In addition, in this letter, we assume that the NS envelope
is composed of heavy elements (e.g., Fe, Co, Ni) with little
accretion of light elements (e.g., H, He, C, O) from the atmosphere;
in this case, the gap activity becomes most active
\citep{hiro13}.
We adopt the canonical NS mass of $1.4 M_\odot$,
and the magnetic dipole moment of $3.2 \times 10^{30} \mbox{G cm}^3$.
In this case, he NS radius becomes $11.6$~km and
the PC magnetic field strength does $4.1 \times 10^{12}$~G.
To examine young pulsar emissions,
we adopt $P=54$~ms and $\dot{P}= 2.6 \times 10^{-13} \mbox{s s}^{-1}$,
which correspond to the NS age of $3$~kyr if the spin down is
due to the magnetic dipole radiation.
To compute the flux, we adopt the distance of $1$~kpc.

\section{3-D Vacuum outer gap model}
\label{sec:vac_3D}
\begin{figure}
 \epsscale{1.0}
 \includegraphics[angle=-90,scale=0.55]{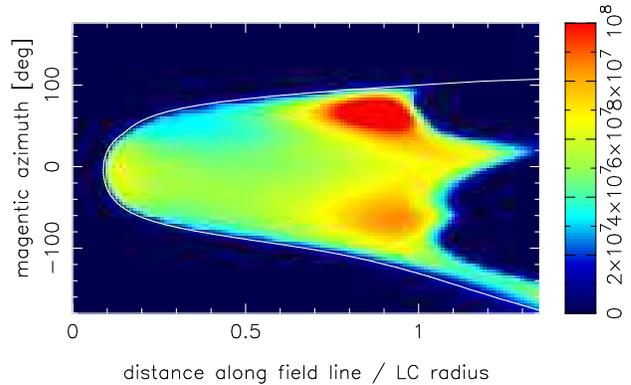}
\caption{
Acceleration electric field in the 3-D Vacuum OG model.
The maximum value of $E_\parallel$ 
[$\mbox{V m}^{-1}$] in the trans-field direction, 
is projected on the last-open-field-line surface 
at each ($s$,$\varphi_\ast$). \ 
\label{fig:Ell_vac}
}
\end{figure}

Let us first examine a vacuum OG in the 3-D pulsar magnetosphere,
by solving the Poisson equation~(\ref{eq:Poisson_1}) under $\rho=0$,
and by assuming $h_{\rm m}=0.08$, which is typical for 
young pulsars around $3$~kyr (within the vacuum OG model).
Since the Poisson equation is a second-order partial differential equation,
and since the gap is transversely thin (i.e., $h_{\rm m} \ll 1$),
$E_\parallel$ distributes quadratically in the trans-field direction, 
and maximizes at the middle height, $h=0.5 h_{\rm m}=0.04$.
We plot this maximum value on the last-open-field-line surface,
($s$,$\varphi_\ast$) in figure~\ref{fig:Ell_vac}.
In this vacuum OG model, the inner (i.e., starward) boundary
is located at the null-charge surface,
whose intersection with the last-open-field-line surface is
indicated by the white solid curve in figure~\ref{fig:Ell_vac}.

It follows that $E_\parallel$ peaks in the higher altitudes
($0.7<s/\rlc<1.0$),
particularly in the leading side ($45^\circ < \varphi_\ast < 75^\circ$).
This is because the GJ charge density per magnetic flux tube
has a greater gradient there compared to the
lower altitudes or in the trailing side,
as indicated by figure~1 of \citet{hiro14}. 
This result forms a striking contrast to the standard OG models,
which extends the 2-D solution of $E_\parallel$ on the poloidal plane
(i.e., at $\varphi_\ast=0$) into the toroidal direction 
(i.e., to $\varphi_\ast \ne 0$ regions).
It means that we must solve the Poisson equation fully three dimensionally
even in the vacuum case.

Using this $E_\parallel$,
we can solve the Boltzmann equations for $e^\pm$'s,
and the emissivity distribution in the 3-D magnetosphere.
Note that in a vacuum OG model, 
the Poisson equation is solved
separately from the Boltzmann equations or the radiative transfer equations.
The particles are accelerated up to the Lorentz factors
$\gamma \sim 3 \times 10^7$
and efficiently emit GeV photons via the curvature process
mainly within the OG, and less efficiently up-scatter 
the magnetospheric IR-UV photons into TeV after escaping from the OG,
where the IR-UV photons are mostly emitted by the secondary pairs
created outside the OG via the synchrotron process.

\begin{figure}
 \epsscale{1.0}
 \includegraphics[angle=-90,scale=0.45]{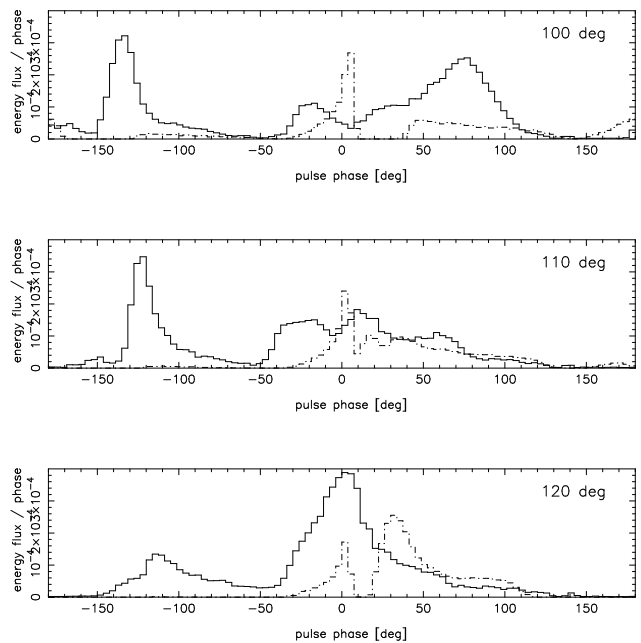}
\caption{
Gamma-ray pulse profiles above 91~MeV 
at three discrete viewing angles for the 3-D vacuum OG model.
The thick curves represent the outward gamma-ray fluxes, while
the thin ones do the inward fluxes.
Photons are emitted not only within the light cylinder,
but also outside of it by the positrons escaped from the gap.
The ordinate ranges from $0$ to 
$4 \times 10^{-4} \mbox{MeV s}^{-1} \mbox{ cm}^{-2} \mbox{ deg}^{-1}$.
\label{fig:res_2b}
}
\end{figure}

The resultant light curves are plotted in figure~\ref{fig:res_2b}
for the three discrete viewing angles with respect to the rotation axis, 
$\zeta=100^\circ$, $110^\circ$, and $120^\circ$.
The solid lines represent the pulse profile of the
outward-propagating $\gamma$-rays (emitted by positrons),
while the dashed ones do that of the inward $\gamma$-rays 
(emitted by electrons).
Note that the vacuum OG model predicts the detection
of the inward emission from the southern OG in addition to 
the conventional outward emission from the northern OG.
This forms a contrast to the two-pole caustic/SG model,
which predicts the detection of 
only the outward emissions from the both poles,
because the leptonic flux is outwardly uni-directional,
and because sufficient emissivity is assumed 
below the null-charge surface.
Since $E_\parallel$ is stronger in the leading side,
the leading peak (P1) tends to be stronger than the trailing peak (P2) 
at many observers' viewing angles, $\zeta$.

Before escaping from the gap, 
typical inward-migrating electrons run $0.3 \rlc$, 
while typical outward-migrating positrons run $0.7 \rlc$.
As a result, outward flux becomes only a few times stronger than the
inward flux. 
Therefore, the light curve in figure~\ref{fig:res_2b} 
generally exhibits 
more than two peaks in a single NS rotation, 
which contradicts with the majority of gamma-ray observations.

\section{3-D Non-vacuum outer gap model}
\label{sec:nonvac_3D}
Let us next consider the screening effect
due to the separating motion of the created pairs in the gap.
We solve equations (43)--(55) in \citet{hiro13}
under the boundary conditions that $e^-$'s or $e^+$'s
do not enter across either the inner or the outer boundaries.
Pairs are mainly created when the inward curvature $\gamma$-rays
collide with the outward thermal X-rays emitted from the NS surface.
The created pairs in the gap are separated to screen 
$E_\parallel$ to a small amplitude so that the pairs can be
marginally separated.
In this case, the real charge density has the same spatial
gradient as the GJ charge density \citep{gol69}
along the magnetic field,
as indicated by figure 5 of \citet{hiro06a}.
We neglect the magnetic-field deformation due to the magnetospheric 
currents, adopting the same magnetic field geometry as 
in section \ref{sec:vac_3D}.

\begin{figure}
 \includegraphics[angle=-90,scale=0.55]{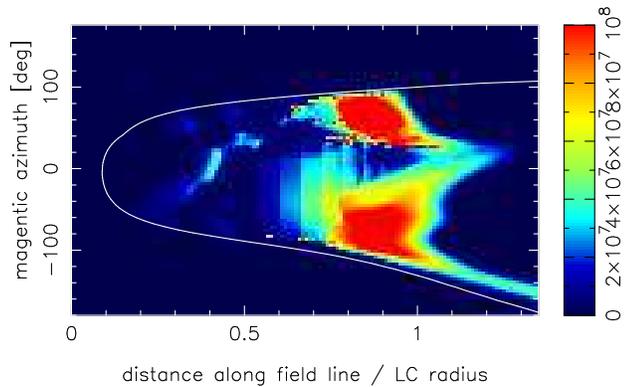}
\caption{
Same figure as figure~\ref{fig:Ell_vac},
but the 3-D non-vacuum OG solution is plotted.
The color code is common with figure ~\ref{fig:Ell_vac}.
%
\label{fig:res_3a}
}
\end{figure}

Because of this screening effect,
$E_\parallel$ becomes very weak
in the middle and lower altitudes, $s<0.77 \rlc$,
as figure~\ref{fig:res_3a} shows. 
It is also found that the regions that emit photons in P1 and P2 phases
(i.e., in $50^\circ < \varphi_\ast < 80^\circ$ and
 $-105^\circ < \varphi_\ast < 45^\circ$ in fig.~\ref{fig:res_3a})
have greater $E_\parallel$ than other regions.
The gap trans-field thickness becomes
$0.11 < h_{\rm m} < 0.13$ 
in most portions of the gap.

\begin{figure}
 \includegraphics[angle=-90,scale=0.45]{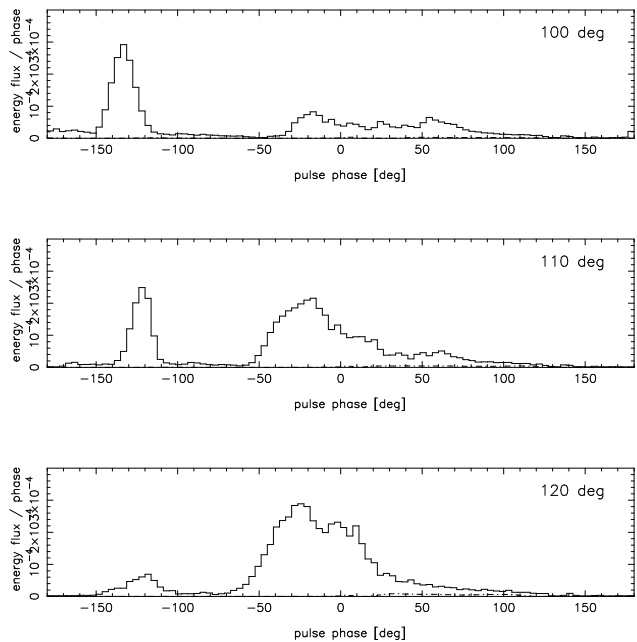}
\caption{
Same figure as figure~\ref{fig:res_2b},
but the light curves for the 3-D non-vacuum OG solution is plotted.
The ordinate range is same with figure~\ref{fig:res_2b}.
Note that only the outward flux (solid curve) appears
in a non-vacuum OG model.
\label{fig:res_3e}
}
\end{figure}

As a result of this $E_\parallel$ screening,
the outward photon flux dominates
the inward one, as demonstrated by the light curves 
in figure~\ref{fig:res_3e}.
This is because the pairs are mostly created in the
middle or the lower altitudes, $s<0.77 \rlc$,
which indicates that the positrons experience an efficient 
acceleration in the strong $E_\parallel$ region in the higher altitudes
while the electrons do not.
Therefore, the light curve is dominated by the outward photons,
which are emitted from the northern OG into the southern hemisphere,
and tends to exhibit 
a double-peak pulse profile for a wide range of $\zeta$.

\begin{figure}
 \includegraphics[angle=0,scale=0.60]{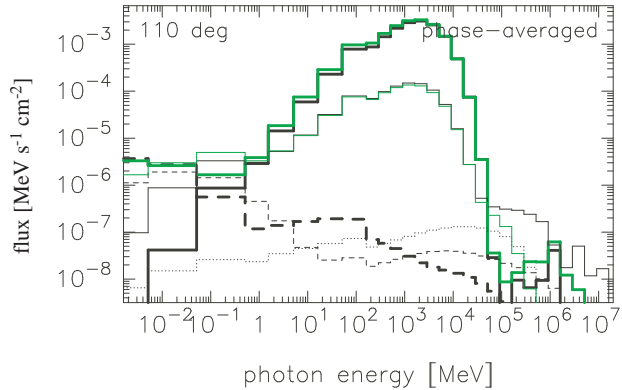}
\caption{
Non-vacuum, 3-D outer-gap solution:
Phase-averaged spectrum at $\zeta=110^\circ$.
The thick lines represent the energy flux of outwardly emitted photons,
while the thin ones do that of inward photons.
The black solid, dashed, and dotted curves denote
the spectrum of the primary, secondary, and tertiary photons,
while the green solid ones do the final spectrum after absorption. 
%
\label{fig:res_3b}
}
\end{figure}

The expected phase-averaged spectrum 
is plotted for $\zeta=110^\circ$ in figure~~\ref{fig:res_3b}.
For comparison, we plot the spectrum of the
outward and inward photons as the thick and thin curves, respectively. 
It is also confirmed from the figure that the gamma-ray flux is
predominantly outward.
As a result of the superposition of the curvature emission
from different places with varying $E_\parallel$,
between 0.16~GeV and 1.6~GeV, the $\nu F_\nu$ spectrum becomes 
a power-law with index $0.68$,
which is consistent with the Fermi observations of young pulsars. 
Since $E_\parallel$ depends not only on ($s$,$\varphi_\ast$),
but also quadratically on $h$, it is essential to
consider the superposition of the emission spectra along
different magnetic field lines.
The intrinsic luminosity of the magnetospheric emission
is $1.7 \times 10^{36} \mbox{ ergs s}^{-1}$ above 160~MeV,
while it is $1.3 \times 10^{35} \mbox{ ergs s}^{-1}$ below 160~MeV.
The heated PC luminosity is $3.3 \times 10^{32} \mbox{ ergs s}^{-1}$.
For comparison, the spin-down luminosity is 
$4.4 \times 10^{37} \mbox{ ergs s}^{-1}$,
and the cooling NS X-ray luminosity is 
$1.2 \times 10^{33} \mbox{ ergs s}^{-1}$.
The heated PC flux becomes greater than the 
magnetospheric X-ray flux if $\zeta < 45^\circ$ or $\zeta > 135^\circ$.

\section{Discussion}
\label{sec:disc}
In summary, we have numerically examined the pulsar outer gaps,
by solving the set of the Poisson equation,
the $e^\pm$ Boltzmann equations,
and the radiative transfer equation from 0.005~eV to 20~TeV.
Applying the method to a young pulsar with $3$~kyr age,
we find that the acceleration electric field $E_\parallel$ is
substantially screened by the separating motion of the
created pairs,
and that the $\gamma$-ray flux becomes predominantly outward
due to the localization of $E_\parallel$ in the higher altitudes.
To reproduce the observed double-peak light curves,
it is essential to solve the outer gap three-dimensionally,
taking account of this screening effect.

As figure~\ref{fig:res_3e} indicates,
the trailing peak has a long tail until the rotational phase of 
$\sim 100$~degrees.
This is due to the emission from the trailing-most side of the
magnetosphere ($\varphi_\ast<-90^\circ$) 
from a very high altitudes ($\rlc < s < 3 \rlc$).
Since the actual strength and direction of such emissions 
depend on the magnetic-field configuration near the LC,
it could be possible to constrain the magnetic field configuration there.
Figure~\ref{fig:res_3e} also shows that there is a strong emission component
around $\varphi_\ast \sim -20^\circ$ (i.e., before P2).
This component is suppressed, if we consider a very thin OG 
(e.g., $h_{\rm m} < 0.05$), as suggested in the standard OG or SG models.
However, if we solve the set of Maxwell-Boltzmann 
equations, we obtain $h_{\rm m} \sim 0.12$ and 
the broad light curves as presented.
It means that we cannot still reproduce the observed flux
and sharp pulses simultaneously by the current particle accelerator models.

Let us briefly discuss an implication when the minimal cooling scenario 
with a heavy element envelope breaks down.
If the NS envelope contains light elements with mass greatly exceeding 
$10^{-16} M_\odot$, the higher NS surface temperature \citep{page04}
leads to a reduction of $h_{\rm m}$ and hence the OG luminosity.
On the contrary, if the cooling is dominated by neutrino emission
via the direct Urca process, 
the resultant rapid cooling (in the initial $\sim 100$~years) 
\citep{negreiros14,coelo14}
will lead to an increase of $h_{\rm m}$ and hence the luminosity.
In either case, we expect that
$E_\parallel$ is localized in the higher altitudes  
in the same way as demonstrated in this letter,
by virtue of the strong negative-feedback effects in the OG electrodynamics
\citep{hiro06b}

Since the optical depth for photon-photon pair creation is
around unity, the TeV photons created via the
synchrotron-self-Compton process cannot be easily absorbed
by the magnetospheric IR-UV photons,
as indicated by the dashed and thick solid curves in
figure~\ref{fig:res_3b}.
It suggests that we may expect relatively strong pulsed emissions
around TeV from the pulsars of which 
inverse-Compton and photon-photon-absorption optical depths
are around unity,  
which is typical for young pulsars with ages around several thousand years.
We will discuss this possibility in a separate paper.

\acknowledgments

The author is indebted to Drs. K.~S. Cheng, J.~Takata, and D.~F. Torres 
for valuable discussion.
This work is supported by 
 the Theoretical Institute for
 Advanced Research in Astrophysics (TIARA) operated under 
 Academia Sinica, and
 the Formosa Program between National Science Council  
 in Taiwan and Consejo Superior de Investigaciones Cientificas
 in Spain administered through grant number 
 NSC100-2923-M-007-001-MY3.

\end{document}